# Dispatch and Primary Frequency Control with Electrochemical Storage: a System-wise Verification

Yihui Zuo, Fabrizio Sossan, Mokhtar Bozorg and Mario Paolone
Distributed Electrical System Laboratory
École Polytechnique Fédérale de Lausanne, Lausanne, Switzerland
Email: yihui.zuo@epfl.ch

*Abstract*—Uncertainty levels in forecasting of renewable generation and demand are known to affect the amount of reserve required to operate the power grid with a given level of reliability. In this paper, we quantify the effects on the system reserve and reliability, due to the local dispatch of stochastic demand and renewable generation. The analysis is performed considering the model of the IEEE 39-bus system, with detailed dynamic models of conventional generation, wind generation, demand and an under-frequency load shedding mechanism. The analysis compares to cases: the base case, where renewable generation and demand power are stochastic and the power reserve is provided by conventional generation, against the case where the operation of traditionally stochastic resources is dispatched according to pre-established dispatch plans thanks to controlling local flexibility. Simulations reproduce the post-contingency dynamic behavior of the grid due to outages of generators. The contingencies are selected to trigger under frequency load shedding mechanisms, hence to demonstrate the different levels of system operation reliability for the two case studies. Simulation results show that dispatching traditionally stochastic generation scores better regarding to expected energy not served, producing an increase of the system reliability.

*Index Terms*--Storage, Primary Frequency Control, Dispatch

## I. INTRODUCTION

The displacement of conventional generation in favor of production from stochastic renewable resources leads to increased fast power system reserve requirements, calling for the support of distributed energy resources for fast regulation (e.g., [1], [2]). Battery energy storage systems (BESSs) are envisaged to play a key role in providing fast reserve thanks to their large ramping capabilities. Compared to conventional generation, BESSs are characterized by much smaller power rating and, therefore, should be deployed in a large number to support power system operations effectively.

A major concern arising from the large-scale deployment of storage units is the coordination of the operation of all BESSs, a complex problem due to the large number of decision variables to handle. The works [3]-[5] describe a scheduling and control framework, called *dispatchable feeder*, to dispatch the power flows of distribution systems at their grid connection point by controlling local resources, like energy storage systems and flexible demand. Thanks to achieving dispatched-by-design operation of traditionally stochastic power distribution grids, the dispatchable feeder's operational paradigm inherently reduces power system reserve requirements with reduced complexity because the control objective is local. In the rest of this paper, we will refer to this operational paradigm as "dispatched-by-design" operation to denote that active power flows of distribution systems, traditionally stochastic, are now deterministic and dispatched according to a pre-established dispatch plans.

This paper focuses on modeling and quantifying the effects, at the system level, of dispatched-by-design distribution grids. We consider the time-domain detailed model of the IEEE 39-bus system in a scenario with an increased proportion of production from renewable generation. Simulations are based on detailed dynamic models of conventional and renewable generation units, demand and an under-frequency load shedding (UFLS) mechanism. The performance of the dispatched-by-design approach is compared against the base case (i.e. where dispatched-by-design distribution grids are not implemented).

The rest of this paper is organized as: Section II describes in detail all the adopted dynamic simulation models, Section III describes the real-time simulation results of two cases (with and without dispatched-by-design). Section IV concludes the paper by summarizing the obtained results.

## II. SIMULATION MODELS

The study illustrated in this work relies on a detailed dynamic model of the IEEE 39-bus power system. It includes dynamic models of conventional generation, renewable generation, and profiles of wind generation and demand from measurements. This section provides all the details of the models.

### A. Electrical power grid

The considered high voltage electrical grid is the 345 kV IEEE 39-bus benchmark power grid described in [6]. As known, it consists in 39 buses, 19 load buses, and 10 generations buses. Its topology is shown in Fig. 1. We modify the original system to account for a larger penetration of renewable power by adding wind generation for a total installed capacity of 1.35 GW. The conventional generation mix consists of hydro-power and conventional thermal power plants. The nominal capacity of generation plants and the location of renewable generation is summarized in Table I.

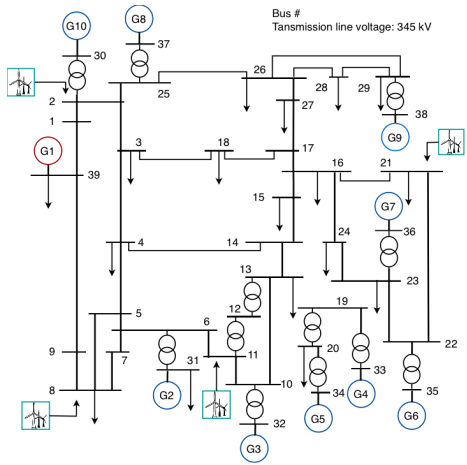

Figure 1.  Diagram of the IEEE 39-bus power grid considered in this work.

TABLE I.  LIST OF GENERATOR UNITS

| Generation plant | Type/ Location | MVA |
|---|---|---|
| G1 | Thermal Plant | 3000 |
| G2–G4, G6–G10 | Hydro Plant | 1000 |
| G5 | Hydro Plant | 520 |
| Wind Farm 1 | bus 2 | 300 |
| Wind Farm 2 | bus 21 | 150 |
| Wind Farm 3 | bus 8 | 400 |
| Wind Farm 4 | bus 11 | 500 |

### B. Conventional generation

Each model of conventional generator consists of a dynamic model of the prime mover (i.e., hydro or thermal, according to the unit type), electric generator (i.e., synchronous machine), the speed governor, DC1A exciter [7] and the associated automatic voltage regulation (AVR). The synchronous machine's model is a sixth-order state-space model from the SimPowerSystem toolbox [8].

The models of the AVR and the prime mover with the associated speed governor are described in the following.

*1) Excitation system and AVR:* We use the IEEE DC1A type excitation system recommended in [7]. Fig. 2 shows the diagram of the excitation system. The $V_{ref}$, $V_g$ and $V_{f0}$ signals are respectively the reference voltage, measured positive sequence voltage and initial field voltage of the generator. Here $K_e = 1$. The compared signal is sent to the AVR, that is represented by a first-order transfer function with a saturation block, as shown in Fig. 2. The parameters are $k_r = 200$, $T_r = 0.001$, $V_{rmin} = 0$, and $V_{rmax} = 12.3$ (p.u.) for steam generators, and $k_r = 200$, $T_r = 0.02$, $V_{rmin} = 0$, and $V_{rmax} = 7.32$ (p.u.) for hydro generators.

*2) Steam turbine governor model:* The model of the steam turbine is based on a complete tandem-compound steam prime mover, equipped with a speed governing system, four-stage steam turbine, and four-mass shaft. The speed governing system of the steam turbine is similar to what proposed in [9]. It consists in a speed governor, speed delay, hydraulic servomotor and governor-controlled valves. The speed-governing system is shown in Fig. 3: the speed reference (SR) signal is set to a constant value as in this work we do not implement automatic generation control; the speed governor is represented by a gain $K_G$ which is the reciprocal of primary frequency droop coefficient; the speed delay is represented by an integrator with time constant $T_{SR} = 0.001$; the hydraulic servomotor is modelled with an integrator with time constant $T_{SM} = 0.15$ and $\dot{C}_{vopen} = 0.1$ and $\dot{C}_{vclose} = -0.1$ as servomotor's speed maximum and minimum limits; the speed of the servomotor is integrated to obtain the position, the maximum and minimum limits of which are $\dot{C}_{vmax} = 4.496$ and $\dot{C}_{vmin} = 0$. The droop coefficient of all the conventional generators is 5%, i.e. $K_G = 20$ for all the steam turbines.

*3) Hydro turbine governor model:* The model of the hydro power plant consists of a non-linear hydraulic turbine model and the associated speed-governing system, which is a servomotor controlled by a PID regulator. The hydraulic turbine model is a non-linear model from the Simulink SimPowerSystem library. The modeling of the hydro turbine's speed-governing system is based on [10]. Fig. 4 shows the speed governor that generates the gate opening signal for the hydraulic turbine. The servomotor is modelled by the first-order system shown in Fig. 4 where $K_a = 3.33$ and $T_a = 0.07$ are the gain and time constant. This model uses the electrical power deviation as droop reference, passing the control signal to a PID controller. The static gain of the governor is equal to the inverse of the permanent droop $R_p$ in the feedback loop. The droop coefficient for all hydro power generation units is 5%.

### C. Load Profile

The three-phase load profiles are voltage and frequency independent where the active and reactive components are inferred by time series data, obtained from a monitoring system based on phasor measurement units (PMUs) installed on the 21-kV grid of the city of Lausanne, Switzerland. The resolution of the time series is 1 second.

### D. Wind Generation

*1) Wind farm modeling:* Each wind farm is modeled as proposed in [11]. It is approximated by multiplying the power output of a detailed model of a single wind turbine to match the total nominal capacity of the farm. The diagram of the overall system in shown in Fig. 5.

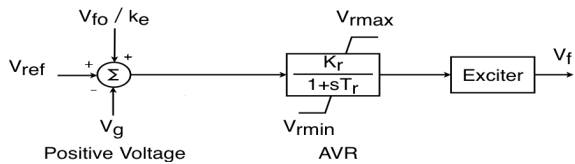

Figure 2.  The IEEE DC1A excitation systme

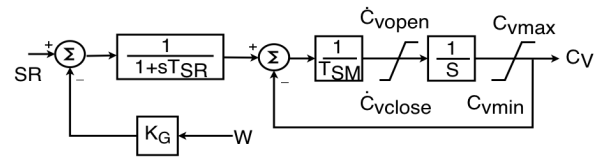

Figure 3.  Diagrm of the speed-governing system of th steam turbine

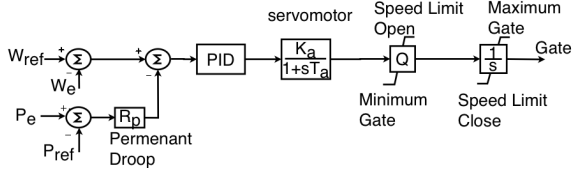

Figure 4. Model of the speed governor of hydro turbine.

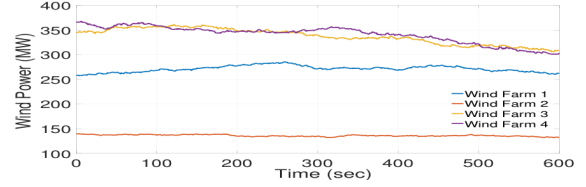

Figure 6. Wind generation profiles adopted in the simulation

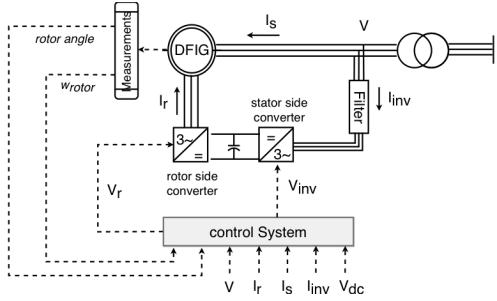

Figure 5. Diagram of the wind farm's model

The model of each wind generator consists of a doubly-fed induction generator (DFIG) and averaged back-to-back converter model from [12]. The detailed aerodynamic model of the wind turbine is neglected as its effect is accounted already in the wind profiles, described in the next paragraph. The back-to-back IGBT voltage-sourced converters (VSCs) are modelled by equivalent voltage sources, which generate the AC voltage averaged over one cycle of the switching frequency. In this averaged converter model, the dynamics resulting from the interaction between the control system and the power system is preserved.

The control system of the back-to-back converter is placed in the DFIG's rotor circuit with the supply-side PWM converter connected to the stator. It uses the terminal voltage $V$, rotor current $I_r$, stator current $I_s$, inverter current $I_{inv}$, DC voltage $V_{dc}$, rotor speed $\omega_{rotor}$ and rotor angle to generate inverter voltage $V_{inv}$ and rotor voltage $V_r$. The inverter voltage $V_{inv}$ is to keep the the DC-link voltage level constant, and the rotor voltage $V_r$ is to enable independent control of the induction machine active and reactive powers.

*2) Wind power profile:* As for load profiles, we rely on time series to model wind generation. Nevertheless, whereas load profiles are from measurements at 1 second resolution, measurements of wind speed or generation at such a high resolution are not publicly available. Therefore, we use measurements at 1~minute resolution from ERCOT (Electric Reliability Council of Texas) [13], which are then re-sampled at 1 second by applying the method described in [14], summarized in the following for the sake of clarity.

Let $x_t$, $t = 1, \dots, T$ denote a time sequence of $T$ aggregated wind power generation values at 1 minute resolution, in per unit. We define a new higher resolution sequence $\omega_{ts}^b$ for $t = 1, \dots, T$, $s = 1, \dots, 60$, at each bus $b$ of the system where wind farms are connected. In other words $\omega_{ts}^b$, is an over sampled version of $x_t$ to be obtained using the following process. The differentiated time series of $\omega_{ts}^b$, is denoted by $\Delta\omega_{ts}^b$ and computed as:

$$\Delta\omega_{ts}^b = \omega_{t,s+1}^b - \omega_{ts}^b, \text{for } t = 1, \dots, T, s = 1, \dots, 60. \quad (1)$$

During each 1 minute time interval $t$, the components of the differentiated time series, $\Delta\omega_{ts}^b, s = 1, \dots, 60$, are sampled from a Gaussian distribution $\mathcal{N}(\overline{\omega_t}, \sigma^2)$, where $\overline{\omega_t} = (x_{t+1} - x_t)/60$ is the average slope of the original wind generation profile, and $\sigma^2$ is estimated based on the statistical characteristics of the aggregated wind generation profiles presented in [14]. Finally, the synthetic profile of wind power generation at 1 second resolution in per unit is:

$$\omega_{ts}^b = x_t + \sum_{s'=1}^{s} \Delta\omega_{ts'}^b. \quad (2)$$

The power generation profile for an aggregated wind farm connected at bus $b$ is:

$$W_{ts}^b = W^b \cdot \omega_{ts}^b. \quad (3)$$

where $W^b$ is the forecast power of the wind farm. Fig. 6 shows the synthetic profiles for 4 wind farms of the modified IEEE 39-bus system.

*E. Transmission Lines*

We adopt the ARTEMiS's distributed parameters line model from Opal-RT [15], which is based on the Bergeron's travelling wave methods used by the electromagnetic transient program (EMTP) [16]. The reason why we use transmission line propagation model is to enable repartition of the simulation computational loads on different cores of the RT simulator (see [15] for further details).

*F. Primary Frequency Control*

All generators, except for the wind turbines, are equipped with droop controllers to implement primary frequency regulation. Since the objective of this paper is to show how the dispatched-by-design strategy impacts on load shedding due to shortages of primary frequency reserve, the generators only implement primary frequency control. As said before, the regulation (droop) coefficients for all the generators are 5%.

*G. Load Shedding*

Under Frequency Load Shedding (UFLS) relays are implemented to curtail loads when the system frequency falls below a certain level. In this paper, UFLS is implemented according to the ENTSOE's (European Network of Transmission System Operators for Electricity) recommendation [17]. Loads are shed proportionally to the frequency drop. The load shedding steps are as following:

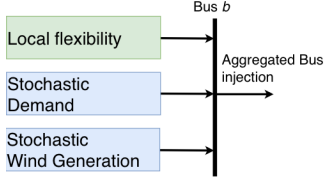

Figure 7. Configuration of dispatched-by-design bus.

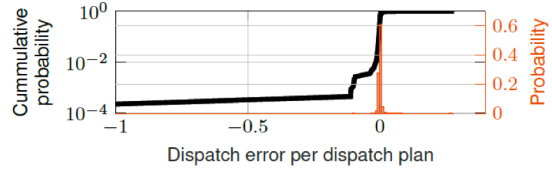

Figure 8. Dispatch error of the dispatchable feeder.

1) from $f_0 - 1.0$ Hz to $f_0 - 1.2$ Hz to shed 5% of load, 2) from $f_0 - 1.2$ Hz to $f_0 - 1.4$ Hz shed 15\% of load, 3) from $f_0 - 1.4$ Hz to $f_0 - 1.6$ Hz shed 25% of load, 4) from $f_0 - 1.6$ to $f_0 - 1.8$ Hz shed 35% of load, 5) from $f_0 - 1.8$ Hz to $f_0 - 2.0$ Hz shed 45% of load, 6) below $f_0 - 2.0$ Hz shed 50% of load.

The UFLS is implemented in all the 19 aggregated loads. It uses PMUs to measure local frequency for the UFLS scheme described above. All the UFLS relays implement a 0.15~second delay to avoid unnecessary UFLS actions. The frequency threshold values for restoring the loads are: $f_0 - 0.25$ Hz for full load restoration, $f_0 - 0.5$ Hz for restoring 95% of the loads, and $f_0 - 0.75$ Hz for 85% restoring.

### H. Dispatch-by-design buses

The dispatched-by-design paradigm achieves to dispatch the operation of stochastic distribution systems by leveraging local flexibility, as for example described in [3]. It is a two-stage process. The first is performed the day before the operation and aims at defining a dispatch plan for the next calendar day while accounting for forecasts of the prosumption, their uncertainty, battery operational constraints and battery discharging/charging demand. The second implements real-time control to achieve a fine tracking of the dispatch plan by controlling flexible resources.

In the simulations proposed in the next Section, power system operations with and without dispatched-by-design are compared. The configuration of a dispatched-by-design bus is shown in Fig. 7. It consists of stochastic demand, stochastic wind generation (if available at that bus), and local flexibility, typically batteries or flexible demand, like in [5]. The aggregated bus injection of a dispatched-by-design bus is the algebraic sum of the three aforementioned components. In case of conventional buses (i.e., non dispatchable), the flexible element is not available and the associated aggregated bus injection is the algebraic sum of stochastic demand and wind generation (if available), only. The operation of the two kinds of buses are explained in the following.

The operation point of the system (i.e., generators set-point and the amount of allocated reserve for primary frequency control) are assumed to be optimized considering the baseline trajectories, like if they are point predictions used in a unit commitment problem. For each case, the stochastic component is sampled from two cases described below.

*1) Case A—without dispatched-by-design:* The wind power $\omega_{ts}^b$ is obtained as presented in section II-D2. The demand power $L_{ts}^b$ at a given bus $b$ is:

$$L_{ts}^b = L^b \cdot l_{ts}^b, \quad (4)$$

where $L^b$ is the forecasted load at bus $b$, and $l_{ts}^b$ is the load profile at 1 second resolution, in per unit, obtained as presented in section II-C.

*2) Case B—with dispatch-by-design:* We assume that the controller We assume that the controllers of BESSs do not provide any frequency and voltage support, therefore, in this case, we neglect their dynamic response.

We define $B_{ts}^{b*}$ as the power injection of the battery at bus $b$ that compensates for the difference between the net power realization ($W_{ts}^b - L_{ts}^b$) and the net scheduled power ($W^b - L^b$):

$$B_{ts}^{b*} = (W^b - L^b) - (W_{ts}^b - L_{ts}^b). \quad (5)$$

However, in practical cases, the performance of dispatched-by-design distribution systems might not be ideal, due to, e.g., outages of battery systems or wrong forecasts. In order to model potential failures of the dispatched-by-design strategy, we model the imperfect battery injection by adding a stochastic component to it:

$$B_{ts}^b = B_{ts}^{b*}(1 + \epsilon_{ts}). \quad (6)$$

where $\epsilon_{ts}$ is sampled from the cumulative probability distribution (CDF) shown in Fig. 8. The CDF is estimated from actual statistics on the operation of the experimental configuration described in [3] considering 16 days of data, from February 6 to February 12, 2017 and from May 1 to May 9, 2016.

## III. RESULTS

### A. Simulation Setup

The power network and simulation models are implemented in MATLAB Simulink and executed on an OPAL-RT real-time digital simulator. The advantage is two-fold: (i) It provides a precise quantification of the computational requirements of the simulation, a key element considering the high computational burden due to extended size of the simulated system. (ii) It allows real-time control operation, which enables the test prototyping of multiple time-scales control strategies (e.g. hourly based commitment with minutes-based real-time optimization).

### B. Case Studies and Scenarios

To evaluate the impact of the dispatched-by-design architecture we compare two cases:

- Case A: power system without dispatched-by-design.
- Case B: power system with dispatched-by-design.

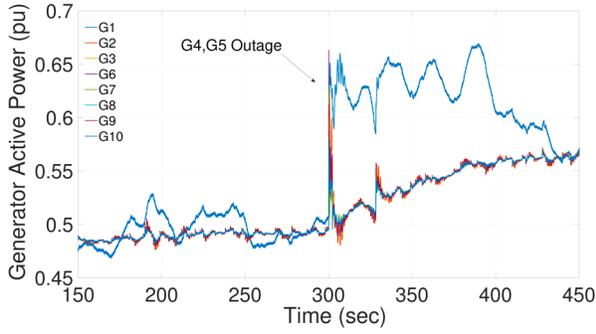

Figure 9. Generators active power scenario S1A

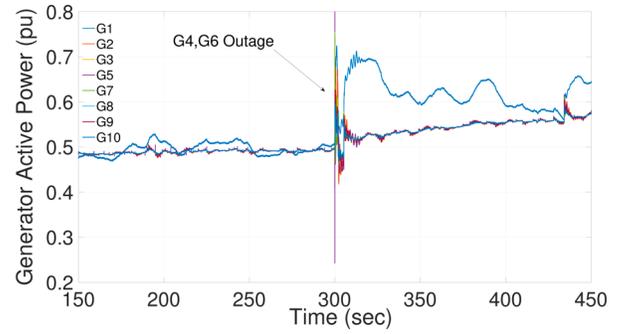

Figure 12. Generator active power scenario S2A

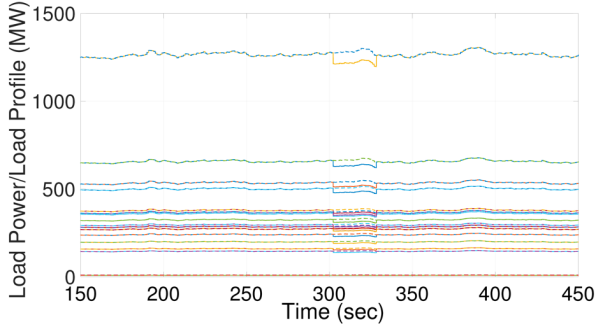

Figure 10. Load shedding for the scenario S1A

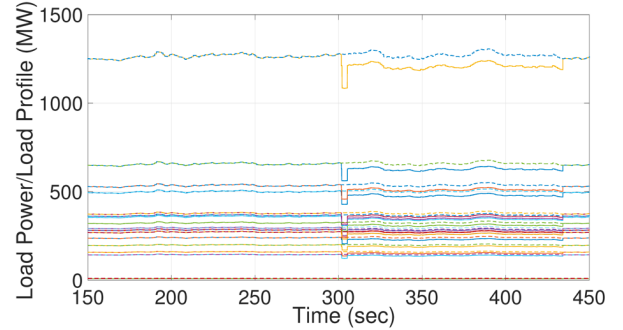

Figure 13. Load shedding for the scenario S2A

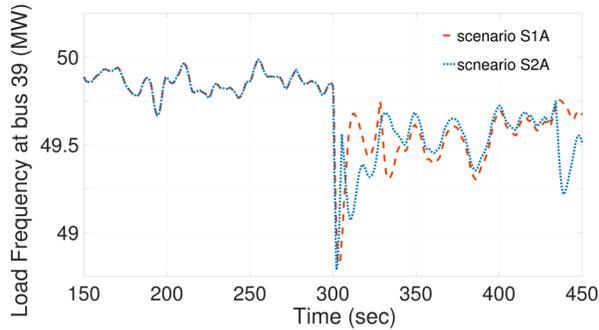

Figure 11. Frequency at bus 39 for the scenario S1A and S2A

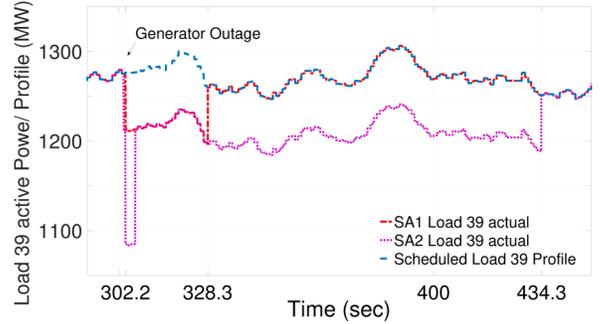

Figure 14. Load at 39 bus scenario S1A and S2A

To have enough power outage to activate UFLS, the selected contingency cases require to trip two generators. We select G4, G5/ G4, G6 as combined outages of generators because, in realistic situation (e.g. generator protection misoperation), it is possible that the closely connected generators would be tripped simultaneously. Therefore, for each case, we consider the following two contingencies:

- Scenario 1: simultaneous outages of G4 and G5.
- Scenario 2: simultaneous outages of G4 and G6.

## C. Results and Discussion

Figures 9-14 show the simulation results of Case A. At time 300~seconds, the contingency happens. Fig. 9 and Fig. 12 show the corresponding active power outputs of the remaining connected generators under contingency 1 and contingency 2. All of the active power outputs increase in order to compensate for the power imbalance. Fig. 10 and Fig. 13 show the actual and expected demand of all the loads. As visible, 5% of the demand is shed in order to reduce the under-frequency in Scenario 1, while in Scenario 2 after a short time of 15% load shedding, it recovered to 5% load shedding. Figures 11 and 14 show the frequency and load active power at bus 39 for both S1A and S2A. As illustrated in Figure 14, the drop of load is due to UFLS being activated in both scenarios after the contingencies (at 300 seconds). In S1A, the load-shedding is triggered 2.2 seconds after the contingency and lasts for 26.1 seconds. In S2A, the load-shedding is triggered at 1.6 seconds after the contingency and lasts for 132.4 seconds. The results for the same two contingencies, but for the dispatched-by-design case, are shown in figures 15-20. Statistics about the load shedding events are summarized in Table II. We use 3 metrics to evaluate the impact of load shedding: maximum load shedding, disruption duration and total EENS. From Table II,

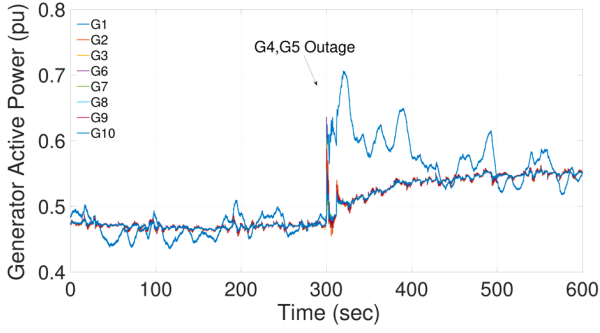

Figure 15. Generators active power scenario S1B

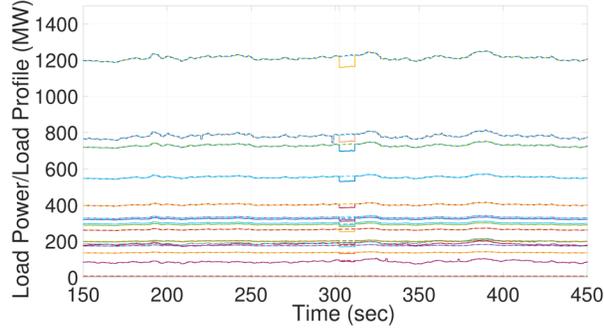

Figure 16. Load shedding for the scenario S1B

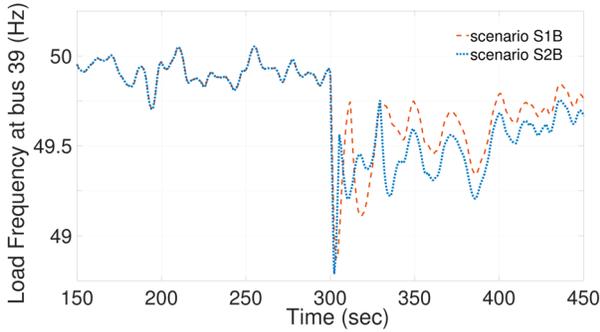

Figure 17. Frequency at bus 39 for the scenario S1B and S2B

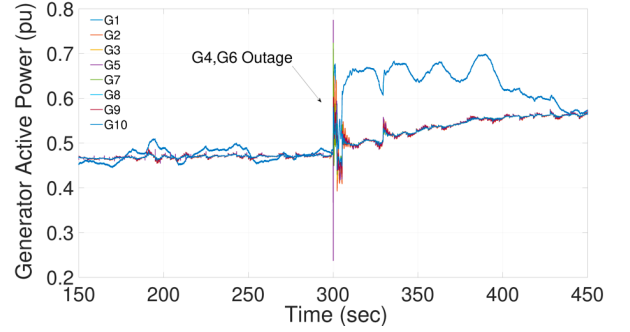

Figure 18. Generator active power scenario S2B

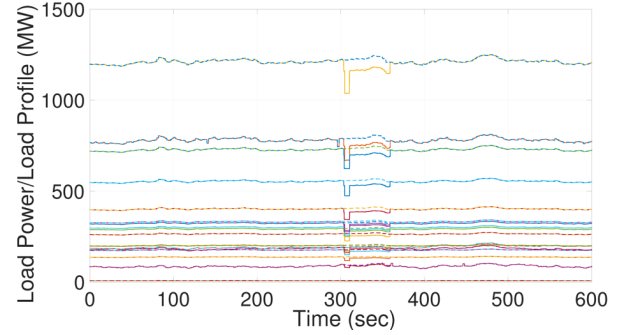

Figure 19. Load shedding for the scenario S2B

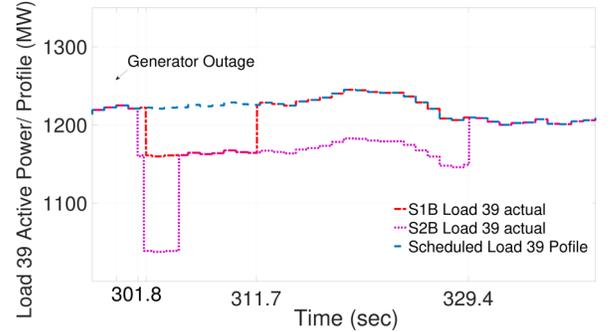

Figure 20. Load at 39 bus scenario S1B and S2B

we can see that: i) load shedding rate $R_{ls}$, load shedding duration $T_{ls}$, and EENS $\Delta E$ of Scenario 2 are larger than Scenario 1 due to the larger amount of tripped generation; ii) by comparing load shedding duration in Scenario 1, $T_{ls}^{S1B}$ is 0.8 second shorter than $T_{ls}^{S1A}$; iii) by comparing load shedding duration in Scenario 2, $T_{ls}^{S2B} = 27.9$ seconds is 78.9% shorter than that $T_{ls}^{S2A} = 132.4$ seconds; iv) by comparing EENS in Scenario 1, $\Delta E_{S1A} = 2.403$ MWh is around 3 times larger than $\Delta E_{S1B} = 0.814$ MWh; v) by comparing EENS in Scenario 2, $\Delta E_{S2A} = 12.829$ MWh is more than 3 times larger than $\Delta E_{S2B} = 3.066$ MWh.

## IV. CONSLUSION

This work assessed how uncertainty levels in forecasts of demand and renewable generation affect the reliability of power systems during real-time operations. The study is conducted by considering the IEEE 39-bus system, implementing detailed dynamic models of conventional generation (i.e., thermal and hydro), wind generation, power electronics, and load shedding mechanisms. Models are specifically formulated to be implemented in a dedicated real-time digital simulator to handle the high computational burden of simulations. With respect to the original IEEE test system, an increased proportion of wind generation is implemented to achieve 20% of wind production during operations. Two case studies are considered and compared: nowadays situation, where aggregated nodal injections of distribution grids and renewable generation are stochastic, against the case where traditional stochastic power flows are dispatched. This is done by emerging the "dispatched-by-design" paradigm from the literature, which achieves to dispatch stochastic resources according to pre-established dispatch plans by leveraging downstream flexibility. Simulation results show that the dispatched-by-design paradigm: (i) decreases the UFLS activating duration, especially for worst contingency in terms of power loss; (ii) reduces the EENS up to a factor 3 during contingencies, improving the system reliability during sudden losses of generation.

TABLE II. SIMULATIONS STATISTIC RESULTS

|  |  | Max Load Shedding $R_{ls}$ | Load Shedding Duration $T_{ls}$ | EENS $\Delta E$ |
|---|---|---|---|---|
| Case A | S1 | 5% | 26.1 s | 2.403 MWh |
|  | S2 | 15% | 132.4 s | 12.829 MWh |
| Case B | S1 | 5% | 25.3 s | 0.814 MWh |
|  | S2 | 15% | 27.9 s | 3.066 MWh |